# Modeling the Biophysical Effects in a Carbon Beam Delivery Line using Monte Carlo Simulation

**Ilsung Cho, SeungHoonYoo, Sungho Cho, Eun Ho Kim,**

**Yongkeun Song, Jae-ik Shin and Won-Gyun Jung***

*Division of Heavy-ion Clinical researches*

*Korea Institute of Radiological & Medical Sciences, Seoul 01812*

Relative biological effectiveness (RBE) plays an important role in designing a uniform dose response for ion beam therapy. In this study the biological effectiveness of a carbon ion beam delivery system was investigated using Monte Carlo simulation. A carbon ion beam delivery line was designed for the Korea Heavy Ion Medical Accelerator (KHIMA) project. The GEANT4 simulation tool kit was used to simulate carbon beam transporting into media. An incident energy carbon ion beam in the range between 220 MeV/u and 290 MeV/u was chosen to generate secondary particles. The microdosimetric-kinetic (MK) model is applied to describe the RBE of 10% survival in human salivary gland (HSG) cells. The RBE weighted dose was estimated as a function of the penetrating depth of the water phantom along the incident beam direction. A biologically photon-equivalent Spread Out Bragg Peak (SOBP) was designed using the RBE weighted absorbed dose. Finally, the RBE of mixed beams was predicted as a function of the water phantom depth.

# I. INTRODUCTION

Carbon ion beams have a potential benefit for the treatment of deep tumors because of their high dose conformity, high relative biological effectiveness (RBE) characteristics, and low oxygen enhancement ratio (OER) [1-2]. Their clinical effectiveness and safety during cancer treatment have been demonstrated by several international institutions for more than 20 years [3].

Carbon ion beams are classified as high linear energy transfer (LET) radiation, which has a higher cell killing effect than that of a photon beam [4]. However, the increasing RBE of carbon beams in the vicinity of the Bragg peak region makes it more complex to utilize carbon ion beams for cancer treatment. Historically, a 10% survival rate in human salivary gland (HSG) tumor cell has been the adopted standard for carbon ion therapy [5-6].

Several biological models have been proposed to account for the biological effectiveness of carbon beams. The theory of dual radiation action (TDRA) proposed by Keller and Rossi is one model to explain the biological effectiveness of ionizing radiation (such as LET) using microdosimetry [7]. Scholz et al. proposed the local effect model (LEM) on the statistical framework [8-9]. The LEM has been adopted in the planning of treatment using carbon ion therapy in German institutions [10] and the Centro Nazionale di Adroterapia Oncologicain (CNAO) in Italy [11]. In the context of TDRA, the microdosimetric-kinetic (MK) model was proposed by Hawkins [12-13]. The MK model is based on the stochastic energy deposition in a micro-scale subcellular structure referred to as a domain. In the MK model, the cell survival fraction of any kind of radiation can be predicted from the microscopic energy deposition in domains. The MK model based approach has been investigated in National Institute of Radiological Sciences (NIRS) in Japan. The MK model was modified by revising saturation correction term which accounts for the decrement of the biological effectiveness due to overkill effect in high LET region [14]. Kase et al. showed that the modified MK model can predict the biological effectiveness from the microdosimetric measurement [14]. Sato et al. evaluated the biological effect of carbon ion beams using the modified MK model for therapeutic carbon ion beams [15]. Inaniwa et al. developed the modified MK model to calculate the biological effectiveness in mixed radiation fields of scanned carbon ion beams [16]. Both the LEM and the modified MK model have been successfully adopted in the clinic for the scanned carbon beam.

The modified MK model shows the ability of a combined theoretical and experimental approach to account for biological effectiveness given that the modified MK model requires the measurement of microdosimetric values [14]. Nevertheless, the application of the modified MK model is not practicable for institutions under development because of the requirement for measured microdosimetric quantity. Recently, Djamel et al. showed that the application of the MK model gives comparable calculation result with the modified MK model for HSG cells and V79 cells in terms of the biological effectiveness [17]. Thus, the MK model has been applied in this study in order to investigate the biophysical characteristics of the designed beam delivery line in the Korea Heavy Ion Medical Accelerator (KHIMA).

The purpose of the study was to establish the calculation framework for the biological effectiveness of carbon ion therapy as part of the KHIMA project. Monte Carlo simulation is an effective way to design new beam delivery lines. The calculation method for determining biological effectiveness is a key component in carbon ion therapy. In this study, a calculation method for determining the biological effectiveness of the KHIMA carbon ion beam delivery line is proposed. This calculation method uses Monte Carlo simulation combined with the MK model.

## II. Material and Methods

**Monte Carlo modeling of a beam delivery line**

The KHIMA project at the Korea Institute of Radiological and Medical Sciences (KIRAMS) has focused on the development of a synchrotron-based accelerator facility with proton and carbon ion sources for the purpose of particle therapy for cancer treatment [18-19]. The layout of the accelerator facility is shown in Fig. 1. The synchrotron accelerator can accelerate the carbon beam from 110 MeV/u to 430 MeV/u. The desired energy range corresponds to a penetration depth of 3.0 g/cm$^2$ to 31.0 g/cm$^2$..

A horizontal beam delivery line of a scanning beam system was designed for the study. Fig. 2 shows a schematic geometric view of the carbon beam delivery line within the KHIMA building footprint. The distance from the last quadrupole of high energy beam transfer (HEBT) to isocenter is 8.5 m for the horizontal beam line. After the last quadrupole, two pairs of the scanning magnets are located along the beam delivery line. Each pair of scanning magnets bends the incident carbon ion beam to a horizontal or vertical direction. The carbon beam window is located at the end of the beam delivery line. Typically, beam delivery systems have a pair of beam monitoring devices [20]. One is the major counter and the other is the secondary counter. If the output between two beam monitors exceeds 10%, the beam delivery system triggers beam interruption for patient safety. A water phantom surface is arranged at the isocenter. The distance from the carbon beam window to the water phantom surface is 2.5 m.

The GEANT4 (GEometry ANd Tracking) [21] toolkit version 9.6p03 was used to simulate the passage of the carbon beam through the beam delivery system and water phantom. The Reference physics List, QGSP_BIC_EMY was chosen to simulate a carbon beam fragmentation, the cascade phenomenon, and the electromagnetic interaction of primary and secondary charged particles. The QGSP_BIC_EMY was the specially designed Reference Physics List for which high level of accuracy is required [22]. This configuration is recommended physics setting of the *Hadrontherapy* contained in the official GEANT4 code [23].

Millions of carbon ions were generated at the position of a carbon beam window with an incident energy range spanning 220 to 290 MeV/u. The energy step was 5 MeV/u for simulation. All carbon ions were assumed to be uniformly scanned in a 10 cm x 10 cm rectangular area that was perpendicular to the beam incident direction. The beam divergence was not considered in the simulation process because the deposited energy in the water phantom was attributed to primary carbon ions and secondary particles [24].

The generated carbon ions passed through two monitoring boxes and entered a water phantom. The dimension of the water phantom was 30 cm x 30 cm x 60 cm in order to take into account the contribution from all secondary particles during the fragmentation process. The water phantom was voxelized with 0.1 mm of depth direction. The deposited energy for each voxel was calculated for all charged particles. The absorbed dose or the physical dose in the water phantom was obtained by adding the deposited energy passing through each voxel. The microdosimetric variable, LET, was calculated for all voxels.

**The calculation of linear energy transfer**

According to ICRU60, the LET is defined as the mean energy lost by charged particles, $LET_\Delta$, as follows [25].

$$LET_\Delta = \frac{dE_\Delta}{dl} \quad (1)$$

In this equation, $dE_\Delta$ refers to the energy loss of all charged particles due to electronic collisions and the energy transfer less than $\Delta$. $dl$ is the traveling distance. If all possible energy transfers are considered, the $LET_\Delta$ becomes the unrestricted LET. The calculation of unrestricted LET for the incident carbon ion beam introduces another complexity due to energy spread in the water phantom. In this simulation, the local mean of the stopping power for the particle energy spectra at a given depth of the water phantom can be unrestricted LET as done in other study [26].

Generally, the dose averaged LET (LET-dose) plays a key role in accounting for the biological effectiveness [4]. The LET-dose is the average value of the stopping power weighted by the local dose. The LET-dose can be calculated via Monte Carlo simulation. In this study, the LET-dose at a given depth z was implemented as follows [26]:

$$L_d(z) = \frac{\sum_{k=1}^{M} dE_k(dE_k/dl_k)}{\sum_{k=1}^{M} dE_k} \quad (2)$$

In this equation, k is the index of the steps during the simulation process at a given depth z. $dE_k$ is the energy deposited at the k[th] step and $dl_k$ is the particle step length.

**Biophysical modeling and calculation of RBE**

The MK model was used to estimate the biological effectiveness [12-13]. In the MK model, the average number of lethal lesions in the nucleus after irradiation is associated with three parameters, $\alpha_0$, $\gamma$ and $\beta$. Each parameter accounts for the single-track hit and double-track hit event on DNA. The single-track hit damage events are composed of two kinematic parameters, $\alpha_0$ and $\gamma$. $\alpha_0$ is the probability associated with single-track, double-strand break damage. $\beta$ represents the probability associated with double-tracks, double-strand break damage. $\gamma$ represents the probability associated with sub-lethal lesions, where the single-track, single-strand break damage is not repaired after irradiation.

The survival fraction $S$ or the average number of lethal lesions in a nucleus $L_n$ after irradiation with dose D follows a linear quadratic equation.

$$-\ln S = (\alpha_0 + \gamma\beta)D + \beta D^2 = \alpha^* D + \beta D^2 = L_n \quad (3)$$

In this equation, $\alpha_0$ and $\beta$ are the cell survival parameters for photons respectively while $\gamma$ is the single event dose mean specific energy in a domain.

Because the average number of lethal lesions in the nucleus is saturated due to the overkill effect, the cell survival parameter $\alpha^*$ is corrected by considering non-Poisson distribution effect of lethal lesions. According to Hawkins [13], the observed surviving fraction S of the cells exposed to low dose D is as follows:

$$S = (1 - \Phi) + \Phi S_n \quad (4)$$

In this equation, $\phi$ is the fraction of the nuclei that suffer a single event after exposure of the cell and $S_n$ indicates the average number of lethal lesions for single hit to the nucleus. In the MK model, $\Phi$ and $S_n$ can be approximated as follows [13]:

$$\Phi \cong \frac{(\alpha_0+\gamma\beta)D + \beta D^2}{(\alpha_0+\gamma\beta)\gamma_n} \quad (5)$$

$$S_n \cong \exp\{-(\alpha_0 + \gamma\beta)\gamma_n\} \quad (6)$$

In these equations, $\gamma$ represents the single-event dose mean specific energy in a domain and $\gamma_n$ reflects the single-event dose mean specific energy in the nucleus. If one replaces $\Phi$ and $S_n$ respectively by equations (5) and (6) in the equation (4) and take the natural log of $S$ and expand in a Taylor series, the survival parameter $\alpha^*$ is corrected as follows:

$$\alpha^* \cong \frac{1-\exp\{-(\alpha_0+\beta\gamma)\gamma_n\}}{\gamma_n} \quad (7)$$

$\gamma$ and $\gamma_n$ can be approximated as follows respectively [13]:

$$\gamma \cong \frac{0.229}{r_d^2} L_d \quad (8)$$

$$\gamma_n \cong \frac{0.16}{\sigma} L_d \quad (9)$$

In these equations, the radius of a domain has $r_d = 0.45$ μm for the HSG tumor cells while the effective area of the nucleus is $\sigma = 75$ μm². The HSG cell survival two parameters for photon beams are $\alpha_0 = 0.13$ Gy$^{-1}$ and $\beta = 0.05$ Gy$^{-2}$ respectively [14].

In this study, a 10% survival rate for HSG cells was chosen to be end point for tumor cell response after carbon beam delivery. The HSG cell survival parameter $\alpha_i^*$ for a given $i^{th}$ incident can be described by using the microdosimetric quantity LET-dose. The LET-dose can be calculated from the GEANT4 simulation.

The RBE corresponding to a 10% survival rate in HSG cells subjected to a carbon ion beam was calculated using the following formula [2, 4]:

$$RBE_{10} = \frac{D_{10,X-ray}}{D_{10,carbon}} = \frac{2\beta D_{10,X-ray}}{\sqrt{\alpha^{*2}+4\beta\ln(0.1)}-\alpha^*} \quad (10)$$

In this equation, $D_{10,carbon}$ is the 10% survival dose of the carbon beam and $D_{10,X-ray}$ is the 10% survival dose using 200 kVp X-rays. The $D_{10,X-ray}$ is 5.0 Gy [14].

From the $RBE_{10}$ formula above, the RBE can be calculated based on the MK model and Monte Carlo simulation. The photon equivalent dose for a 10% survival rate, the biological dose $D_{bio}$, was obtained by multiplying $RBE_{10}$ by the physical dose $D$:

$$D_{bio} = RBE_{10}\, D \quad (11)$$

**The biological effectiveness of mixed beams**

A Spread-out Bragg peak (SOBP) is a superposition of many Bragg peaks with different penetration depths. Modeling using superposition of Bragg peaks was used. In order to achieve a biologically equivalent tumor response, the biological dose was used instead of the physical dose. The physical dose and $RBE_{10}$ were obtained from Monte Carlo simulation as described in the above paragraph.

The biologically equivalent SOBP was composed by summation of the biological dose curves with a proper weight:

$$SOBP_{bio} = \sum_i w_i D_{bio,i} \quad (12)$$

In this equation, $D_{bio,i}$ is the biological dose for a certain penetration depth, and $w_i$ represents i[th] weight of the i[th] biological dose.

In order to estimate the proper weights, a minimization process was performed using the cost function, $Q^2$ as described below:

$$Q^2 = (SOBP_{flat} - SOBP_{bio})^2 \quad (13)$$

In this equation, $SOBP_{flat}$ is the target SOBP distribution, and $SOBP_{bio}$ is the desired SOBP distribution.

A uniform distribution was taken for the minimization goal so that the tumor response would be biologically equivalent for the target region of the SOBP. The MINUIT [27] and the ROOT [28] analysis frameworks were employed to minimize the cost function, $Q^2$.

The cell survival fraction and the cell survival parameters for mixed beams can be calculated as follows [7]:

$$S_{mix}(D) = \exp(-\alpha^*_{mix}D - \beta_{mix}D^2); \quad (14)$$

$$\alpha^*_{mix}(z) = \sum f_i \alpha^*_i(z) \quad (15)$$

$$\sqrt{\beta_{mix}}(z) = \sum f_i \beta_i(z) \quad (16)$$

In the above equations, $f_i$ is the weight of $i^{th}$ biological dose.

The $RBE_{10}$ for mixed beam was calculated using the mixed cell survival parameters, $\alpha^*_{mix}(z)$ and $\sqrt{\beta_{mix}}(z)$. Finally, the physical SOBP distribution was obtained by dividing the $RBE_{10,mix}$ value by the appropriate depth.

### III. RESULTS AND DISCUSSION

**The Application of the MK Model**

The calculation of the cell survival parameter $\alpha^*$ was performed by using the MK model formulae. The calculation result for $\alpha^*$ value is applied to the experimental data [29]. Fig. 3 shows the results of the cell survival parameter $\alpha^*$ calculation for HSG cells as a function of the LET-dose. The calculated $\alpha^*$ value had a maximum value in the vicinity of the LET-dose at 200 keV/μm. Application of the MK model to the cell survival parameter $\alpha^*$ yielded results consistent with the experimental results. The $\alpha^*$ was used to calculate the $RBE_{10}$ value of HSG cells for biological dose of the designed carbon beam delivery line.

**Carbon ion beam transport**

Fig. 4 shows the simulation results for the monochromatic 290 MeV/u carbon ion beam. The entrance region of the Bragg curve was normalized to unity. The physical dose distribution of 290 MeV/u for the carbon ion beam is shown on the top panel of Fig. 4. The Bragg peak position corresponds to a depth of 16.8 cm in the water phantom.

The LET-dose distribution is shown in the middle of Fig 4. The LET-dose increased steeply in the near Bragg peak and rapidly decreased after the Bragg peak. This behavior was mainly due to the stopping power. Because the LET-dose is the averaged value of the energy-weighted electronic stopping power, the LET-dose followed the Bragg peak behavior. In order to use the LET-dose for $RBE_{10}$, the LET-dose was calculated for every proper depth.

The $RBE_{10}$ for the carbon ion beam is shown in the bottom panel of Fig 4. The $RBE_{10}$ was calculated according to the proper depth using the survival fraction parameters of the HSG cells, $\alpha^*$ and β, and the LET-dose. The maximum value for $RBE_{10}$ reached 3.2. However, two maximum peaks were found on the $RBE_{10}$ distribution. This was mainly due to the saturation correction in the MK model and the LET-dose distribution. The MK model predicted the maximum $RBE_{10}$ to be around 150 keV/μm of the LET-dose value. The LET-dose value increased over 150 keV/μm and reached 300 keV/μm around the Bragg peak region. The $RBE_{10}$ reached the maximum value(3.2) around 150 keV/μm of the LET-dose value, and decreased until the depth corresponds to the Bragg peak point. However, after the Bragg peak region, the LET-dose decreased again and dropped below 150 keV/μm . In contrast, the $RBE_{10}$ reached the maximum value(3.2) again due to the decreasing LET-dose.

The biological dose of 290 MeV/u for the carbon ion beam is overlaid on the physical dose distribution in the top panel of Fig 4. The biological dose is the product of $RBE_{10}$ and the physical dose, by definition. The biological dose distribution had a slightly greater width than the pristine Bragg curve because of the double peaks on the $RBE_{10}$ distribution. The ratio between the entrance region and the Bragg peak region was proportional to the $RBE_{10}$.

**Biological SOBP distribution and $RBE_{10}$ for a mixed Beam**

The left side of Fig.5 shows the biological Bragg curves in the energy range from 220 MeV/u to 290 MeV/u with a 5 MeV/u energy step. The correlation between penetration depth of the incident carbon beam into water-equivalent media and the incident carbon ion beam energy is shown on the right hand side of Fig. 5.

The range of the biological Bragg peaks was 10.5-16.8 cm in water depth. The Bragg curves of corresponding peak depth were chosen from Fig.5. A total of 30 biological Bragg curves were chosen to build a 6 cm biological SOBP distribution. The minimization result is depicted in Fig. 6.

The biological SOBP distribution of the carbon ion beam is shown on the left side of Fig. 7. The biological SOBP was normalized to the center of the SOBP region. The entrance level of the biological SOBP was 0.5. The $RBE_{10,mix}$ for the designed SOBP was calculated using the TDRA [7]. The $RBE_{10,mix}$ is shown on the right side of Fig.6. The $RBE_{10,mix}$ started at 1.2 and reached the maximum value of 3.1. The biological SOBP and the $RBE_{10,mix}$ were consistent with the results of the MK modeling approach used at NIRS [14-15]. The physical dose distribution is calculated by dividing the designed SOBP by the $RBE_{10,mix}$. The designed physical dose can be used to operate a carbon beam delivery system that ensures a uniform dose response in tumor cells.

## V. CONCLUSION

In conclusion, a calculation framework for the biophysical characteristics of a beam delivery line for the KHIMA project has been established. The beam delivery line was designed using the KHIMA building layout. The biological effects of the designed beam delivery line were investigated using Monte Carlo simulation in combination with the MK model. Based on the MK model and Monte Carlo simulation results, a 6 cm biological SOBP distribution corresponding to a 10% survival rate in HSG cell was designed. The RBE for a mixed beam, $RBE_{10,mix}$, was calculated for the KHIMA beam delivery line. The biological effectiveness plays an important role in carbon ion therapy. The designed physical dose and the weight of each physical Bragg curve can serve as the carbon beam modulation parameters. The calculation result of the biological effectiveness will be applied to beam modulation apparatus and finally integrated into treatment planning system.

## ACKNOWLEDGEMENT

This work was supported by the National Research Foundation of Korea (NRF) grant funded by the Korea government's Ministry of Science, ICT and Future Planning (MSIP) (2015001637)

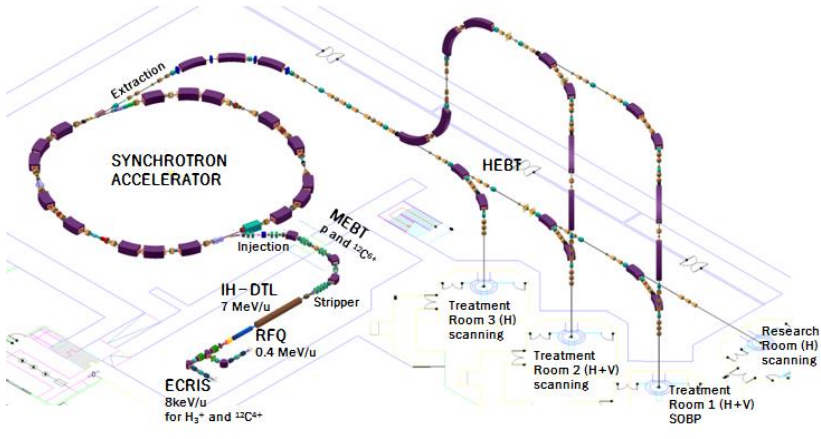

**Fig. 1. Schematic view of the Korea Heavy Ion Medical Accelerator (KHIMA) synchrotron system.**

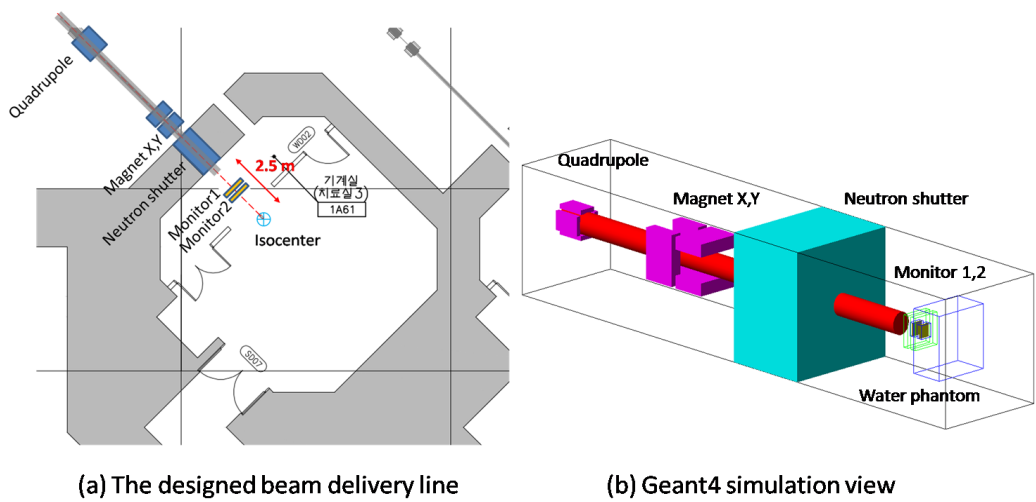

(a) The designed beam delivery line    (b) Geant4 simulation view

**Fig. 2 Schematic view of the designed beam delivery line. The beam delivery line is overlaid on the building footprint (left). GEANT4 simulation view of the designed beam delivery line (right).**

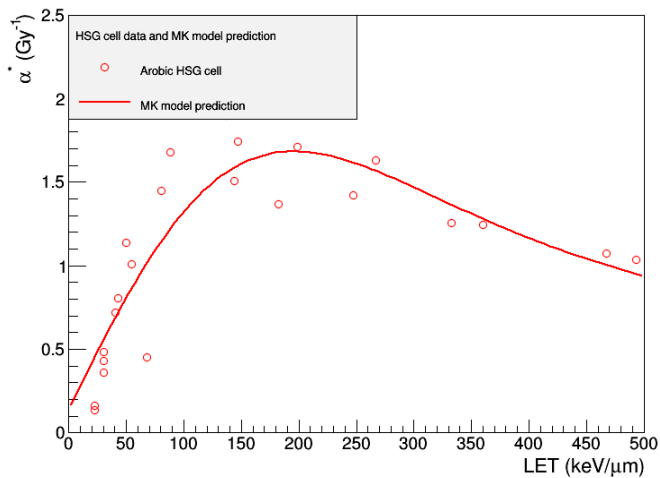

**Fig. 3 The results of the cell survival parameter α* calculations for human salivary gland (HSG) cells. The experimental data are overlaid on the microdosimetric-kinetic (MK) model results.**

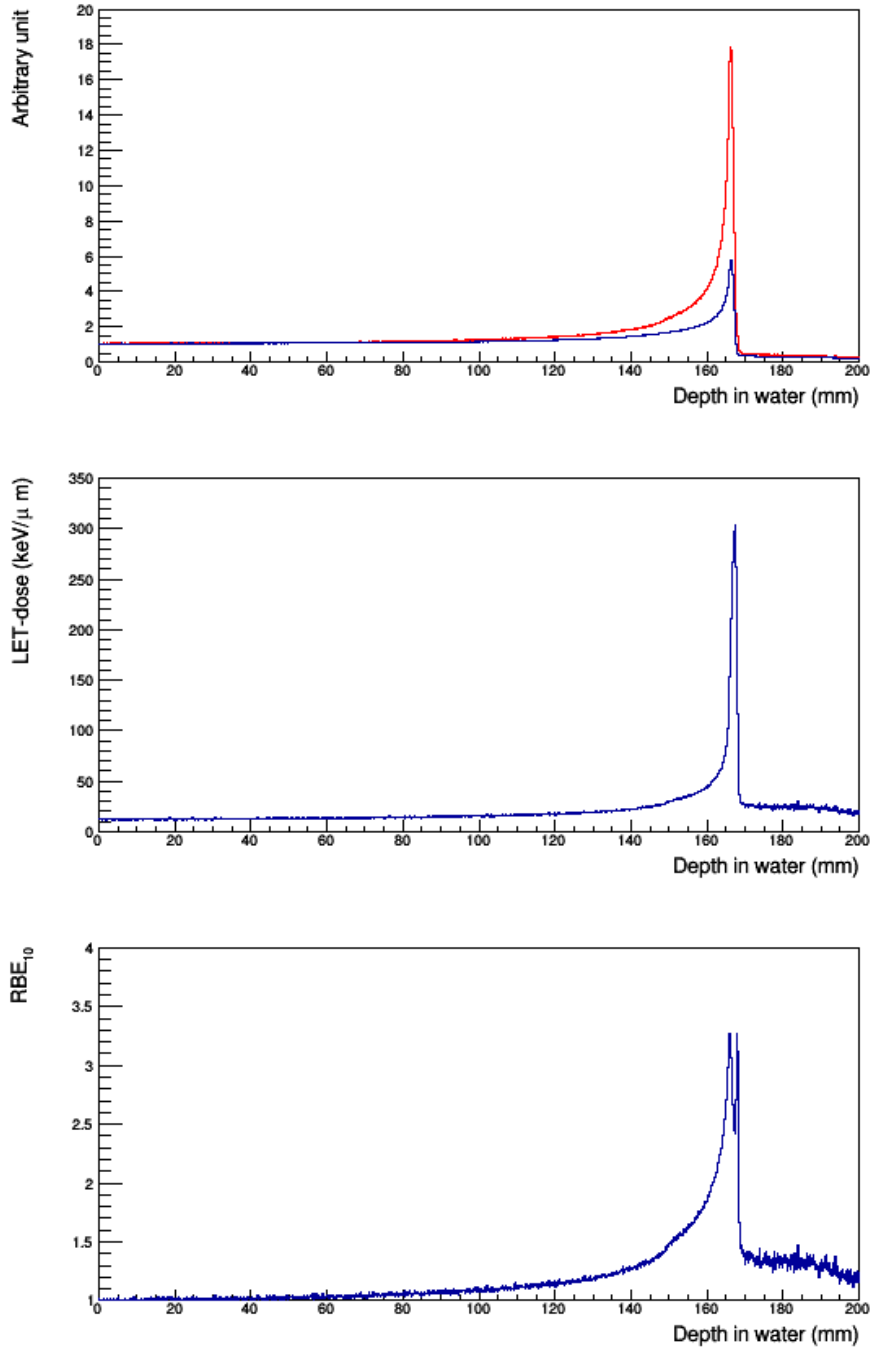

**Fig. 4** The simulation results for the monochromatic 290 MeV/u carbon ion beam. The physical dose (blue) and the biological dose (red) distribution (top). The dose-averaged linear energy transfer (LET) distribution (middle). The relative biological effectiveness (bottom)

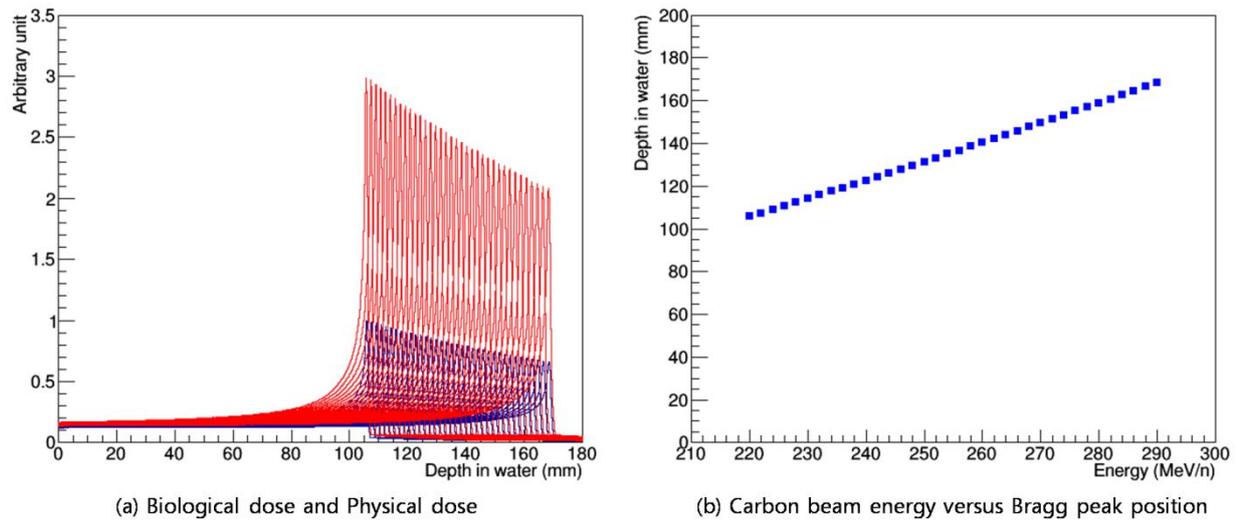

**Fig. 5** The biological Bragg curves in the energy range from 220 MeV/u to 290 MeV/u with a 5 MeV/u energy step (left). The correlation between penetration depth of the incident carbon beam into water and the incident carbon ion beam energy (right)

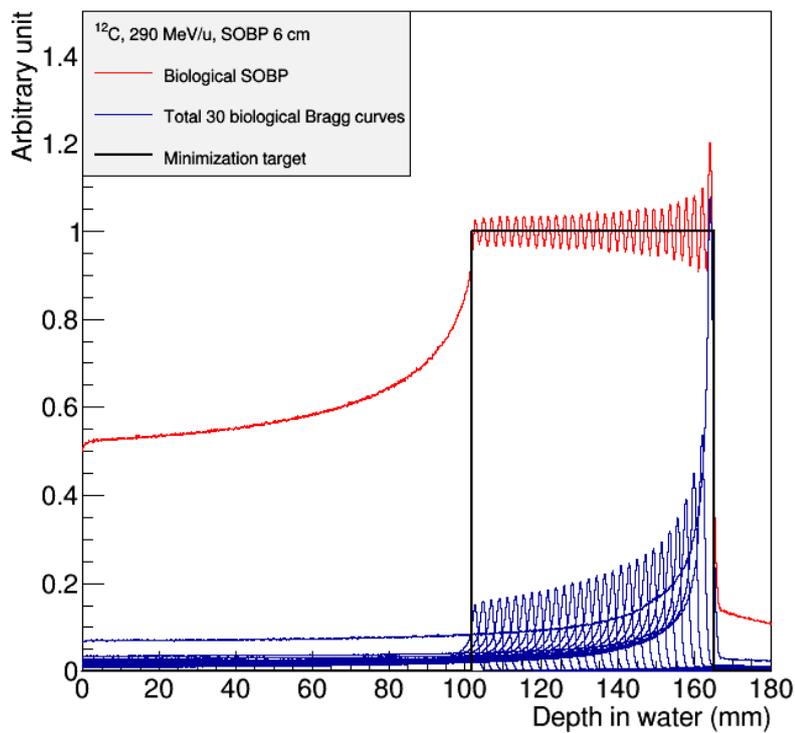

**Fig. 6** The result of Spread-out Bragg peak (SOBP) parametrization. The black line is the minimization target and the red line is the biological SOBP. A total of 30 biological Bragg curves were used for SOBP parametrization (blue line).

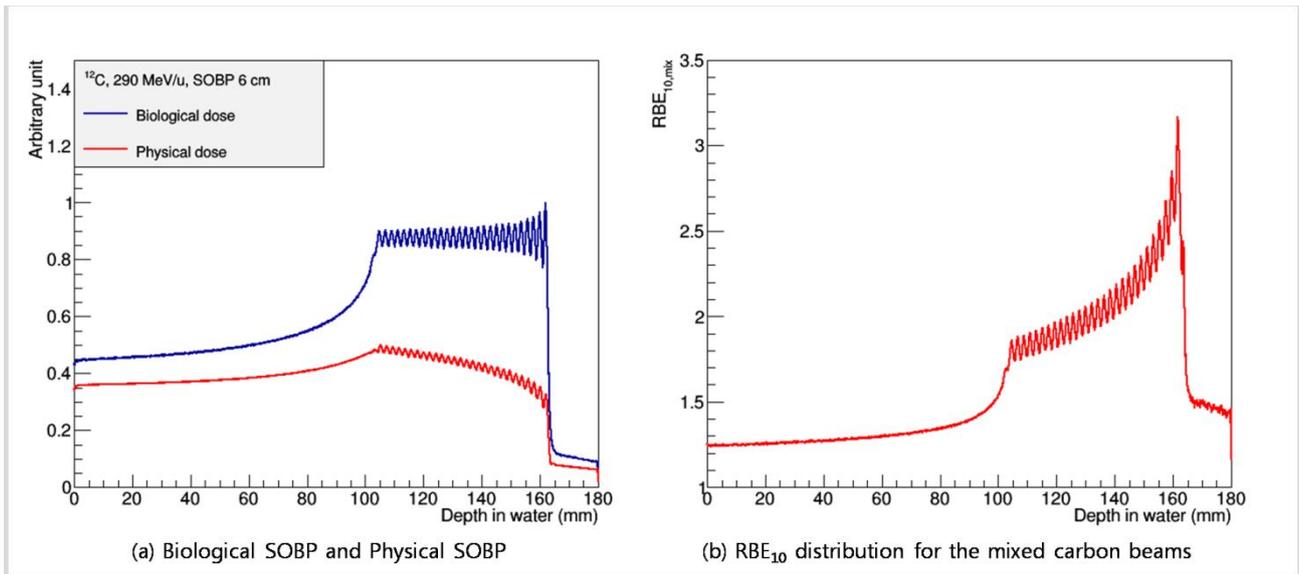

**Fig. 7** A value of 6 cm of the biological Spread-out Bragg peak (SOBP) in water and the corresponding physical dose (left). The calculation result for the relative biological effectiveness for mixed beam corresponding a 10% survival rate in human salivary gland cells (right).